\documentclass[nocompress]{spie}  

\usepackage{amsmath,amsfonts,amssymb}

\usepackage{graphicx}
\usepackage{svg}
\usepackage[colorlinks=true,citecolor=blue]{hyperref}
\usepackage{xspace}
\usepackage{bm}
\usepackage{gensymb}
\usepackage{color}
\usepackage{relsize}
\usepackage{mathtools}
\usepackage{aas_macros}
\usepackage{xcolor}
\usepackage{booktabs}
\usepackage{amsmath}
\usepackage{deluxetable}

\newcommand\arcsec{\mbox{$^{\prime\prime}$}}%
\newcommand\farcs{\mbox{$.\!\!^{\prime\prime}$}}%

\title{Exo-NINJA at Subaru: fiber-fed spectro-imaging of exoplanets and circumstellar disks at R $\sim$ 4000}

\author[a]{M.~El Morsy}
\author[b]{J. Lozi}
\author[b,c,g,h]{O. Guyon}
\author[a,b]{T. Currie}
\author[b]{S. Vievard}
\author[d,e]{J. Bryant}
\author[f]{C. Tokoku}
\author[b]{V. Deo}
\author[b]{K. Ahn}
\author[d,e]{F. Crous}
\author[d,e]{A. Wang}
\author[d,e,]{Z. Sathi}

\affil[a]{Department of Physics and Astronomy, University of Texas at San Antonio, San Antonio, TX 78006, USA}
\affil[b]{National Astronomical Observatory of Japan, Subaru Telescope, 650 North Aohoku Place, Hilo, HI 96720, U.S.A. }
\affil[c]{{Steward Observatory, University of Arizona, Tucson, AZ 85721, USA}}
\affil[d]{Sydney Institute for Astronomy, School of Physics, The University of Sydney, NSW 2006, Australia}
\affil[e]{Astralis Instrumentation Consortium, Physics Building, Physics Road The University of Sydney NSW 2006, Australia}
\affil[f]{National Astronomical Observatory of Japan, 2-21-1 Osawa, Mitaka, Tokyo 181-8588, Japan}
\affil[g]{Astrobiology Center of NINS, 2-21-1 Osawa, Mitaka, Tokyo 181-8588, Japan}
\affil[h]{Wyant College of Optical Sciences, University of Arizona, Tucson, AZ 85721, USA}

\authorinfo{Send correspondence to M. El Morsy (\href{mailto:mona.elmorsy@utsa.edu}{mona.elmorsy@utsa.edu}).}

\begin{document} 
\maketitle

\begin{abstract}

Exo-NINJA will realize nearIR R$\approx$4000 diffraction-limited narrow-field spectro-imaging for characterization of exoplanets and circumstellar disk structures. It uniquely combines mid-R spectroscopy, high throughput, and spatial resolution, in contrast to CHARIS, which does spectro-imaging, and REACH, which is single-point (no spatial resolution).
Exo-NINJA's spectro-imaging at the telescope diffraction limit will characterize exoplanet atmospheres,  detect and map (spatially and spectrally) gas accretion on protoplanets, and also detect exoplanets at small angular separation ($\lambda$/D) from their host star by spectro-astrometry.
Exo-NINJA will link two instruments at the Subaru Telescope using a high-throughput hexagonal multi-mode fiber bundle (hexabundle). The fiber coupling resides between the high contrast imaging system SCExAO, which combines ExAO and coronagraph, and the medium-resolution spectrograph NINJA (R$=$4000 at JHK bands). Exo-NINJA will provide an end-to-end throughput of 20\% compared to the 1.5\% obtained  with REACH.
Exo-NINJA is scheduled for implementation on the Subaru Telescope's NasIR platform in 2025; we will present a concise overview of its future installation, laboratory tests such as the throughput and focal ratio degradation (FRD) performance of optical fiber imaging hexabundles, in the NIR and the trade-offs for fiber choices for the NINJA-SCExAO hexabundle fiber cable, and the expected on sky performance. 

\end{abstract}


\newpage
\section{INTRODUCTION} 
\label{sec:intro}  

Over the past decade, the development and integration of optical fibers into ground-based telescope instruments have revolutionized the field of exoplanet characterization \cite{Currie2023PPVII}. These advancements have primarily focused on optimizing the transfer of exoplanet light to spectrographs, thereby enhancing the precision and depth of observational data. Instruments designed for medium spectral resolution (R = 1000) (e.g. \cite{Petrus2021}) have emerged as particularly effective, striking a balance between the limitations of low-resolution spectrographs, such as CHARIS (R = 20-70)\cite{Groff2016}, which are constrained by their spectral resolution, and high-resolution spectrographs like IRD/REACH (R = 100,000)\cite{kotani2020reach}, which lack spatial resolution.

Medium-resolution spectrographs have enabled significant breakthroughs in exoplanet research. Notably, they have facilitated the detection and abundance determination of molecules in the atmospheres of exoplanets such as HR8799 bc and beta Pictoris b \cite{Barman2015,Konopacky2013,Hoeijmakers2018}.
They can also yield radial velocity for exoplanets \cite{Ruffio2019}.  While these capabilities underscore the importance of medium-resolution instruments in advancing our understanding of exoplanetary atmospheres and dynamics, many such instruments are fed light from facility adaptive optics (AO) systems achieving low Strehl ratios (e.g. SR $\sim$ 0.2-0.5) and thus with modest detection capabilities.

Exo-NINJA is a novel instrument poised to advance exoplanet characterization, uniquely combining medium-resolution spectroscopy, high throughput, and spatial resolution and fed from the leading AO system in the northern hemisphere: the upgraded AO3K facility AO system (first-order correction) and Subaru Coronagraphic Extreme Adaptive Optics project (SCExAO; second-order correction) \cite{Lozi2022,jovanovic2015subaru}. 
Exo-NINJA's capability to perform spectro-imaging at the telescope's diffraction limit behind an extreme AO system will enable comprehensive characterization of exoplanet atmospheres, detailed spatial and spectral mapping of gas accretion onto protoplanets, and the detection of exoplanets at small angular separations from their host stars through spectro-astrometry.

The innovative design of Exo-NINJA involves linking two systems at the Subaru Telescope using a high-throughput hexagonal multi-mode fiber bundle called a hexabundle \cite{2014Bryant} developed at Astralis-USyd at the University of Sydney. The fiber coupling will link SCExAO, which combines extreme adaptive optics (ExAO) and a coronagraph, to the medium-resolution spectrograph NINJA (R=4000 at JHK bands). This setup is anticipated to deliver an end-to-end throughput of 20\%, a significant improvement over the 1.5\% achieved with REACH, thereby enhancing observational efficiency and data quality.

Scheduled for implementation on the Subaru Telescope’s Nasmyth infrared (NasIR) platform in 2025, Exo-NINJA will undergo extensive laboratory testing. These tests will assess the total Exo-NINJA system performance ahead of deployment to the telescope. The hexabundle throughput and FRD have already been characterised. Additionally, the project is exploring the trade-offs involved in selecting fibers for the NINJA-SCExAO hexabundle. The goal is to ensure optimal performance and integration, thereby maximizing the instrument's scientific output.
This paper presents a detailed overview of Exo-NINJA's upcoming installation and expected contributions to exoplanet research. We will discuss the technical and scientific objectives, the laboratory tests, including throughput and FRD performance, and the anticipated on-sky performance. By validating a technical solution for a near-IR fiber integral field unit for NINJA, the proposed work aims to enable a wide range of non-exoplanet observations. The bundle input could be located directly at the telescope’s NasIR focus to enhance throughput for observations that do not require high-contrast capabilities, such as imaging narrow fields within galaxies. We will explore these options in collaboration with the NINJA team, broadening the scope and impact of this innovative instrument.

\section{General overview of Exo-NINJA}
\label{sec:exoninja}
\begin{figure*}
 \centering
 \includegraphics[width=\textwidth]{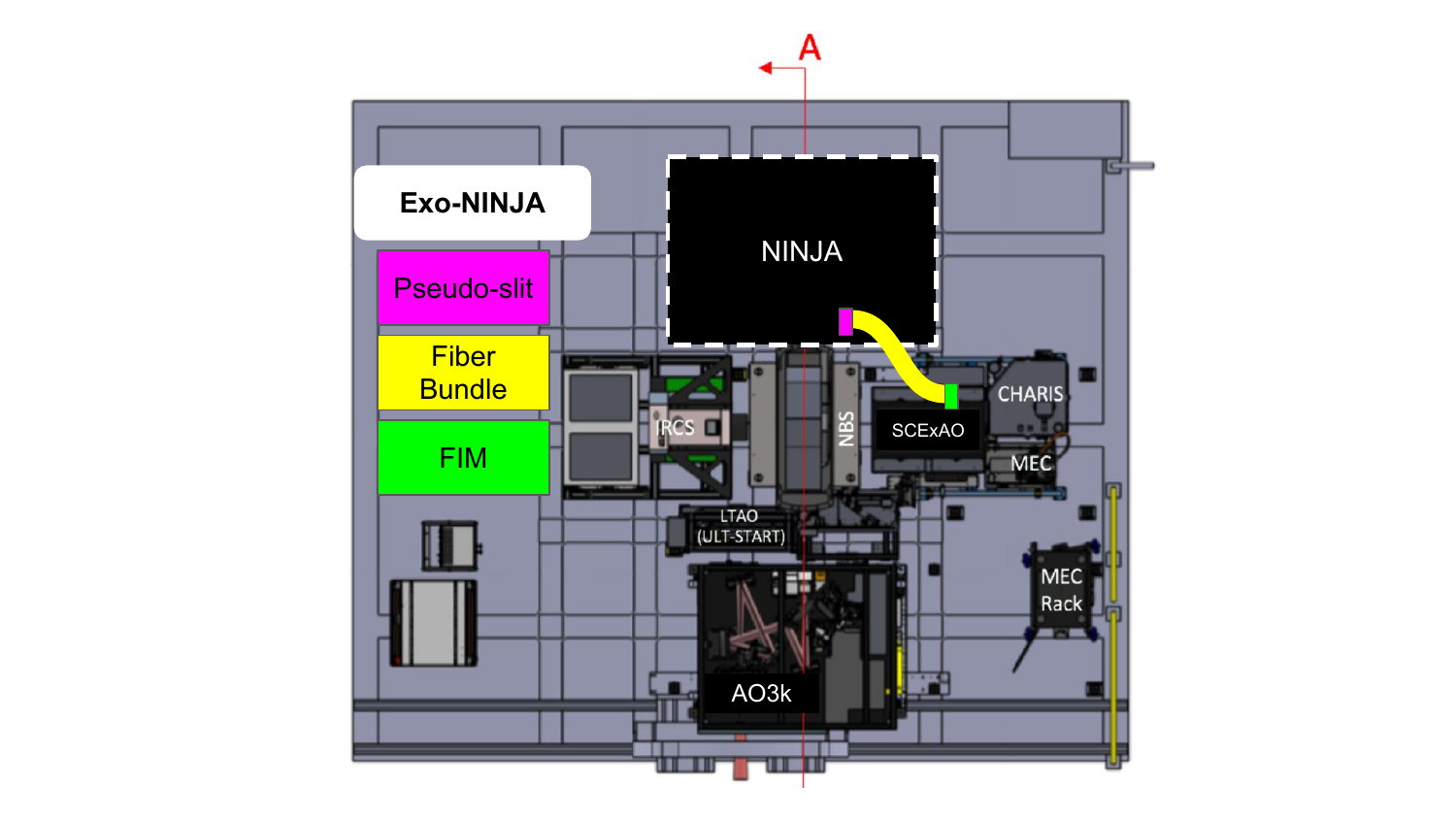}
 \caption{Schematic overview of the future Exo-NINJA implemention at Subaru Telescope. The NINJA spectrograph is connected to SCExAO via a fiber bundle. The green and purple insets indicate the connection points on either side of the coupling.}
 \label{fig:overview}
\end{figure*}

Fig.\ref{fig:overview} displays a schematic of Exo-NINJA as it will be installed at the Subaru Telescope.  Exo-NINJA couples SCExAO with the medium-resolution spectrograph NINJA (R $\sim$ 4000 at the $JHK$ passbands) using a hexagonal multi-mode fiber bundle (hexabundle).   
Exo-NINJA benefits both from the high stability of the point-spread function (PSF) delivered by AO3K+SCExAO and the NINJA spectropgraph, which is optimized for its sensitivity and spectral coverage (0.83-2.5 $\mu m$).

\noindent The Exo-NINJA instrument is comprised of three parts: a fiber-injection module within SCExAO, the NINJA spectrograph, and a fiber bundle connecting SCExAO with NINJA.  We detail each of these three components below.

\subsection{The SCExAO Fiber Injection Module}
\label{sec:SCExAO}
SCExAO is a extreme AO adaptive optics system for the Subaru telescope \cite{jovanovic2015subaru} with a wavelength coverage from 0.6 $\mu m$ to 2.4 $\mu m$, located downstream of Subaru's facility AO system, which is now upgraded to AO3K: a 3000-actuator deformable mirror \cite{Lozi2022} and wavefront sensing capability in the near-IR.   SCExAO has played a leading role in the discovery and characterization of young exoplanets and planet-forming disks, primarily at low spectral resolution with the CHARIS integral field spectrograph (e.g. \cite{Lawson2020,Currie2022,Currie2023Science,Tobin2024,Vincent2024}).    The AO3K upgrade yields extreme AO corrections even prior to starlight being further sharpened by SCExAO, including for optically-faint stars down to $H$ = 8-10.  As a result, the platform yields an improved correction of atmospheric turbulence, a more stable PSF, deeper raw contrasts, and thus better planet detection capabilities for downstream instruments.  

The fiber injection module connected to the MMF is already implemented in SCExAO and validated on the sky by testing the injection of HR 8799 into the fiber. The MMF is placed in the focal plane on a three-axis mount to enable optimized PSF injection. To verify the injection onto the fiber, a tracking camera is used to perform an injection map by scanning the fiber in XY axis, recording the flux measured for each position on the focal plane camera. This enables the construction of an injection map, with a 2D Gaussian fit allowing retrieval of the X and Y sub-pixel positions for which the maximum flux is reached.
In order to maximize the coupling efficiency, the wavefront error must be minimized. For this purpose, the deformable mirror of SCExAO can be utilized to compensate for these errors. We can use the Zernike optimization method described in Delorme et al. \cite{Delorme2021}, which consists of calibrating and reducing the Non-Common Path Aberrations (NCPA) between the imaging and injection paths.

The FIM will play a significant role in the Exo-NINJA project, as it will retrieve the PSF of the target delivered by SCExAO and precisely inject it into the fiber bundle mounted on the three-axis mount. However, to adapt it for SCExAO-NINJA, some optical design options are foreseen and are discussed in the section below.

\subsection{The NINJA Spectrograph}
\label{sec:NINJA}

The NINJA spectrograph, supported by the Japan Society for the Promotion of Science (JSPS), will be installed on the Subaru Telescope NasIR platform and connected to one of the outputs of the Subaru Nasmyth Beam Switcher, which allows multiple instruments to reside on the platform simultaneously and observers to switch between them without craning (installing or removing) the instruments to and from the platform. NINJA will be implemented in different phases, with the first phase taking place in 2025 as part of the ULTIMATE-Subaru Tomography Adaptive optics Research experimenT (ULTIMATE-START) project \cite{masayuki2020}, focusing solely on the near-IR spectrograph. 

NINJA is designed to be a versatile instrument fed by multiple different AO systems, where the Exo-NINJA configuration is but one of the different usages.
Its NIR spectrograph is optimized for the laser tomography adaptive optics (LTAO) system that will be implemented behind the AO3K.
LTAO will enable NINJA to achieve a narrow Field of View (FoV) with diffraction-limited resolution. The first upgrade of LTAO will consist of upgrading the LTAO WFS and implementing four laser guide stars (LGSs).

The Ultimate-START project's second phase aims to develop and implement the ground-layer adaptive optics (GLAO) system, which will cover a FoV of 14x14 arcmin and will consist of off-axis WFSs and an adaptive secondary mirror (ASM) of 924 actuators correcting at a speed of 1kHz. The combination of GLAO and AO3K will be optimized for NINJA. Additionally, the optical spectrograph in NINJA is under development and will be implemented in the second phase of the Ultimate-Subaru project.
Therefore, Exo-NINJA will benefit from the high sensitivity, wavelength coverage, and mid-spectral resolution of NINJA in the near IR.

\begin{figure*}
 \centering
 \includegraphics[width=\textwidth]{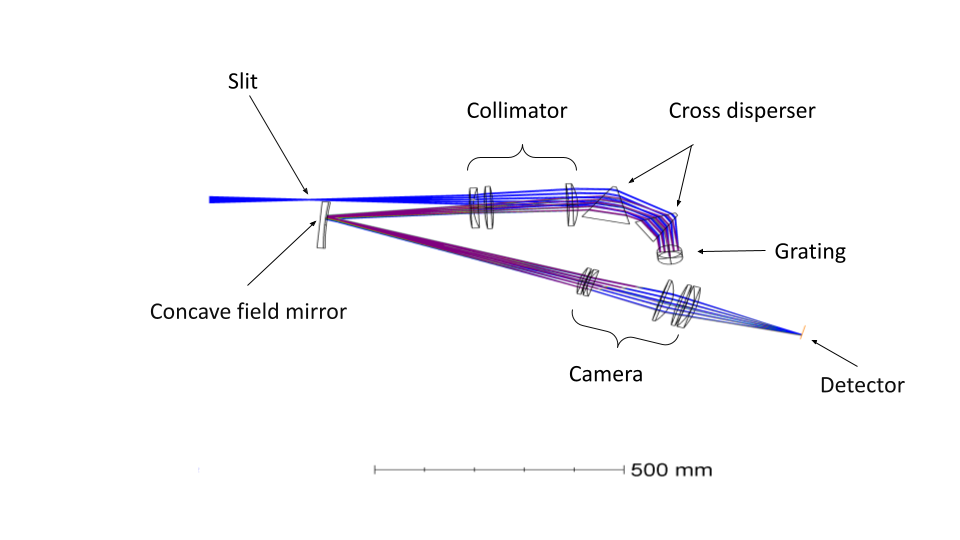}
 \caption{Optical design of NINJA NIR spectrograph.}
 \label{fig:ninja_od}
\end{figure*}

The optical design of NINJA is shown in Fig.\ref{fig:ninja_od} and Table\ref{tab:ninja} summarizes the parameters of the NIR spectrograph. NINJA is based on a white pupil layout design. The pseudo-slit is imaged at the entrance of NINJA's slit, after which the light beam is collimated using lenses and passes through cross-dispersing elements and a grating used in quasi-Littrow configuration. The diffracted light is directed back along the same path as the incident light, with a slight deviation to avoid total overlap. The light beam is then reflected towards the imaging path by a concave field mirror to a detector.

\begin{deluxetable}{lllcllllll}
     \tablewidth{0pt}
     \tabletypesize{\scriptsize}
     \setlength{\tabcolsep}{0.25pt}
     \label{tab:ninja}
    \tablecaption{NINJA Specifications}
    \tablehead{\colhead{Wavelength Range} &  \colhead{F ratio} & \colhead{Slit width,} & \colhead{Slit length} & \colhead{Focal Length} & \colhead{Pupil diameter} & \colhead{Detector}  
    & \colhead{Sampling}\\
     & \colhead{(Input)} & \colhead{Spectral Resolution} & & \colhead{(Collimator)} & &(Format, Pixel Size)} 
    \startdata
0.83--2.5 $\mu$m & 13.9 & 0\farcs{}35, R$\sim$3300 & 5\arcsec{} & 597.7 mm & 43.0 mm & HAWAII-2RG  & 3.3 pixels (0\farcs{}35 slit) \\
 && 0\farcs{}21, R$\sim$5500 & & & & (2048x2048 pixels, 18 $\mu$m)\\
 &&  0\farcs{}50, R$\sim$2310&\\
 &&  0\farcs{}70, R$\sim$1650 &\\
    \enddata
    \end{deluxetable}
    
\subsection{Fiber bundle}
\label{sec:bundle}
The fiber bundle connecting SCExAO to NINJA is a prototype manufactured by Astralis-USyd at the University of Sydney in Australia under contract with Subaru Telescope/NAOJ. The manufactured fiber bundle is optimised from 0.37 $\mu m$ to 2.2 $\mu m$ using Optran WF. A second hexabundle optimised for $>$2.2 $\mu m$ may be a future upgrade. Anti-reflective coatings on input optics, and the output fibre slit with nanostructures will ensure minimal ($\sim$0.1\%) air/glass interface loss across the entire spectral range.

\begin{figure*}
 \centering
 \includegraphics[width=\textwidth]{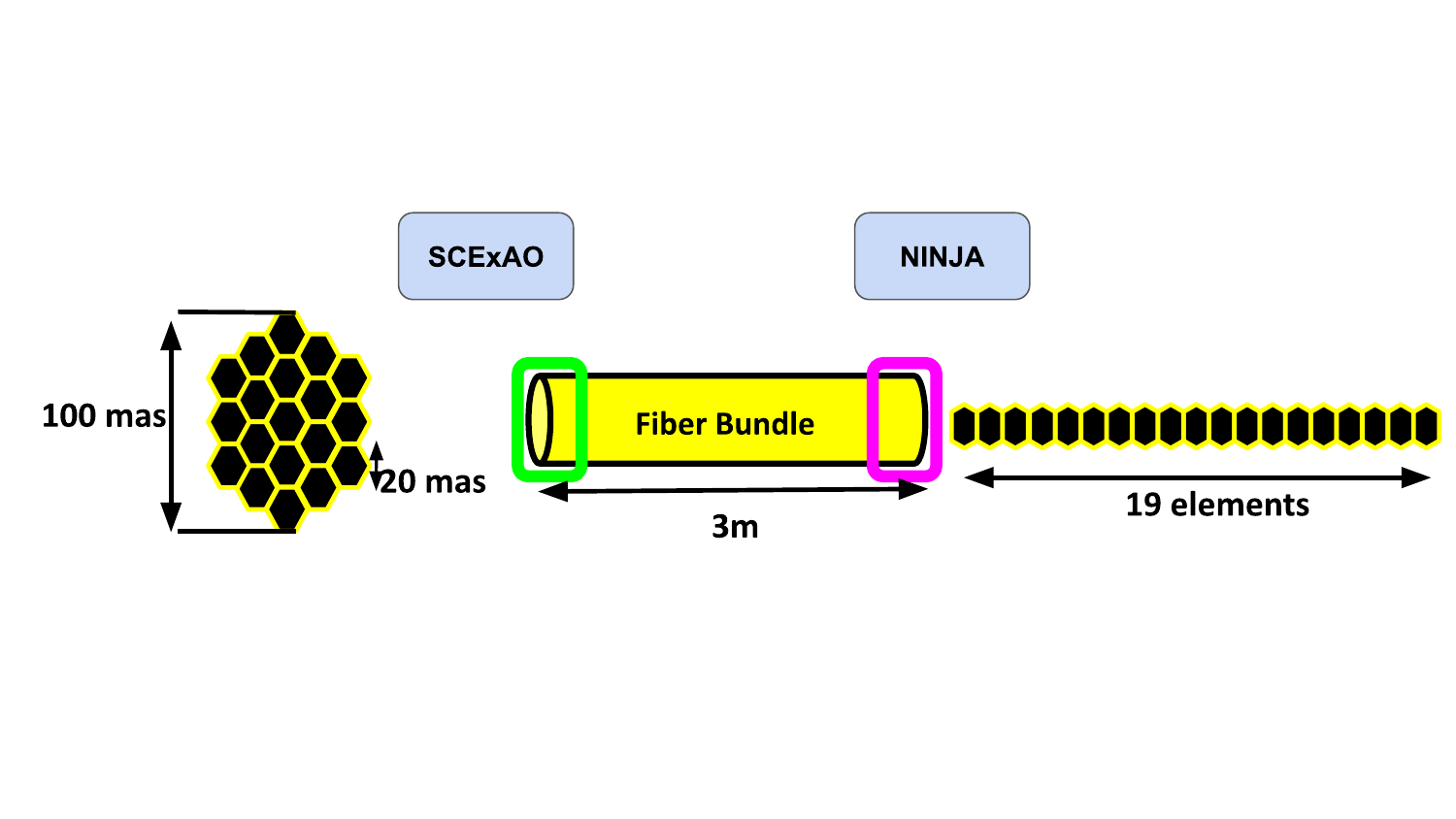}
 \caption{Schematic drawing of the architecture of active zone of the hexagonal fiber bundle . The hexagonal shape of the input fiber bundle optimizes packing density, maximizing the active area available for light capture and transmission.
 The output end of the fibre bundle is 19 individual fibre pigtails that are arranged into a linear slitlet. 
 The purple and green insets represent the respective connection points in SCExAO and NINJA as displayed in Fig\ref{fig:overview}. 
 }
 \label{fig:bundle_drawing}
\end{figure*}

\begin{figure*}
 \centering
 \includegraphics[width=0.8\textwidth]{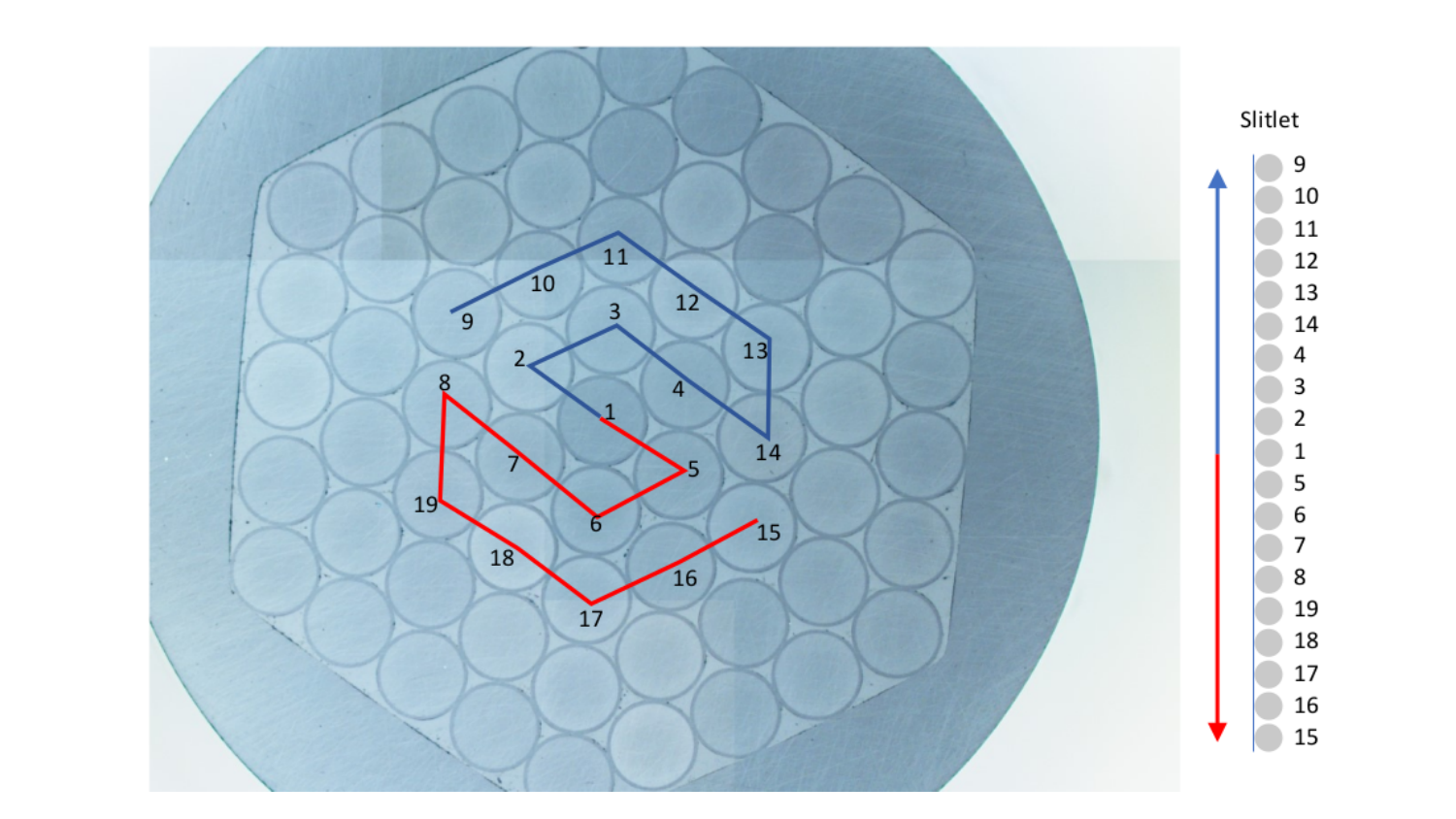}
 \caption{Left: Front view of the illuminated hexabundle on the microscope. The hexagonal pattern represents the active science fibers. Some outer fibers are designated for calibration. The central 19 cores are optimized for optical performance. The diameter of the outer circular capillary is 1.36 mm. Right: Suggested mapping of the hexabundle fibers to the slitlet. Credits: private communication from Julia Bryant.}
 \label{fig:bundle_pic}
\end{figure*}

The fiber bundle input to be implemented on SCExAO as reproduced on Fig. \ref{fig:bundle_drawing} consists of a 2D hexagonal array of several multimode fibers (19 MMF in the active area), each with a 20mas projected on-sky angular diameter. The hexabundles have reduced fibre cladding and are a fused glass bundle which delivers a high fill factor (light into the cores compared to lost between the cores). The prototype hexabundle as displayed in Fig. \ref{fig:bundle_pic} was constructed with a 123$\mu$m outer cladding diameter and a 103$\mu$m core diameter, resulting in a 75\% fill fraction and a measured cross-talk of less than 2\% at 0.8 $\mu$m. A microlens array will be developed for the input face to change the beam speed into the fibre (to optimise FRD) without losing plate scale.

On the NINJA side, the output of the fiber bundle is arranged as a 1D array of 19 MMFs in a pseudo-slit, 2.7mm in length, mounted on a v-groove mount.

Fig.\ref{fig:bundle_pic}, left panel, shows the illuminated fiber input. The hexagonal pattern shows the active zone of the fiber bundle and how the fibres map from the hexabundle to the slit. The central 19 fibers are dedicated to the science path. The outer fibers are dedicated to calibrations and PSF alignment in the active zone.
Fig. \ref{fig:bundle_pic}, right panel, displays a possible rearrangement of the fiber bundle input into a slitlet configuration, which is proposed to minimize the impact of fiber overlapping.

\subsection{Optical design}
\label{sec:opticaldesign}

Single-mode fibers (SMFs) are ideally used at the focal point of telescopes for their spatial filtering properties and efficient light injection into spectrographs. While the use of SMFs has already proven successful in High Dispersion Coronagraphy (HDC) projects, the adoption of multimode fibers bundle (MMFs) could offer improved fill factor due to minimized core-to-core distance and better throughput. However, the advantages of using MMFs are countered by a major drawback, namely focal ratio degradation (FRD), intrinsic to MMF properties. 
FRD in optical fiber systems refers to the distortion of the focal ratio of an incoming light beam as it passes through the fiber, stemming from factors like modal noise, bending-induced losses, and scattering, which impact the output beam profile.

\begin{figure*}
 \centering
 \includegraphics[width=0.6\textwidth]{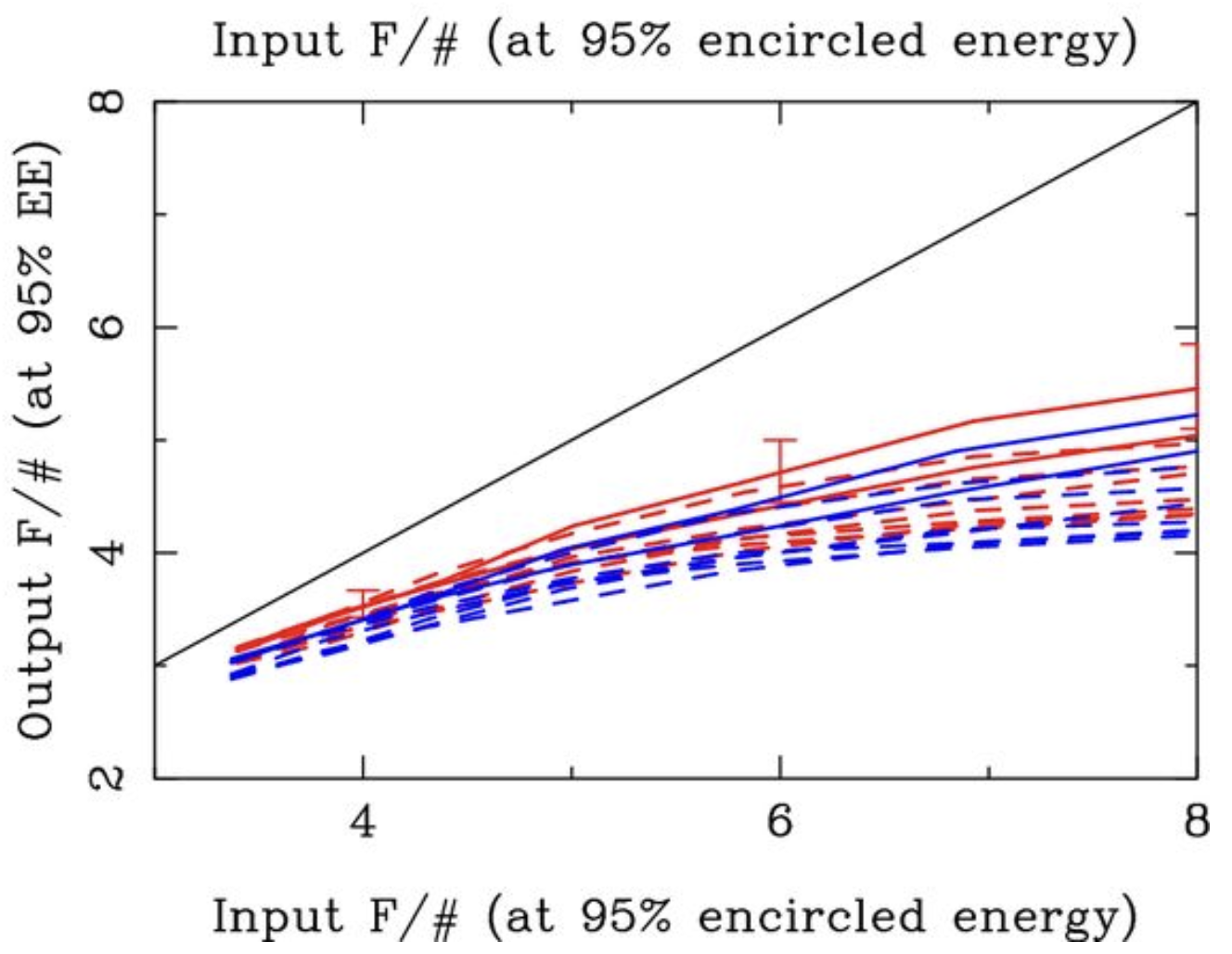}
 \caption{Output f-ratio versus input f-ratio at 95 percent encircled energy. FRD outcomes for the central cores (shown as solid lines) and peripheral cores (represented by dashed lines) in two 61-core hexabundles. The FRD in hexabundles has been shown to be no higher than that in the bare fibre from which hexabundles are made. Reproduced from Bryant et al. 2014\cite{2014Bryant}}.
 \label{fig:FRD}
\end{figure*}

In the case of Exo-NINJA, the initial design proposes the use of a hexabundle. Fig. \ref{fig:FRD} illustrates the output f-ratio as a function of the input f-ratio for a 95\% encircled energy for fibres used to make these bundles, showing the strong F/$\#$ dependence of FRD. 
It is observed that for approximately an F/3 input into the fiber, the FRD is minimized within the fiber. Currently, the input fiber f-ratio on the SCExAO side is F/140. At that F-ratio, the losses from FRD are measured to be 80\%. Therefore, the beam speed must be changed before the injection into the hexabundle.  At the entrance of the spectrograph, the f-ratio must be 14 for optimal spectral resolution and light injection efficiency into the spectrograph. 

\begin{figure*}
 \centering
 \includegraphics[width=0.9\textwidth]{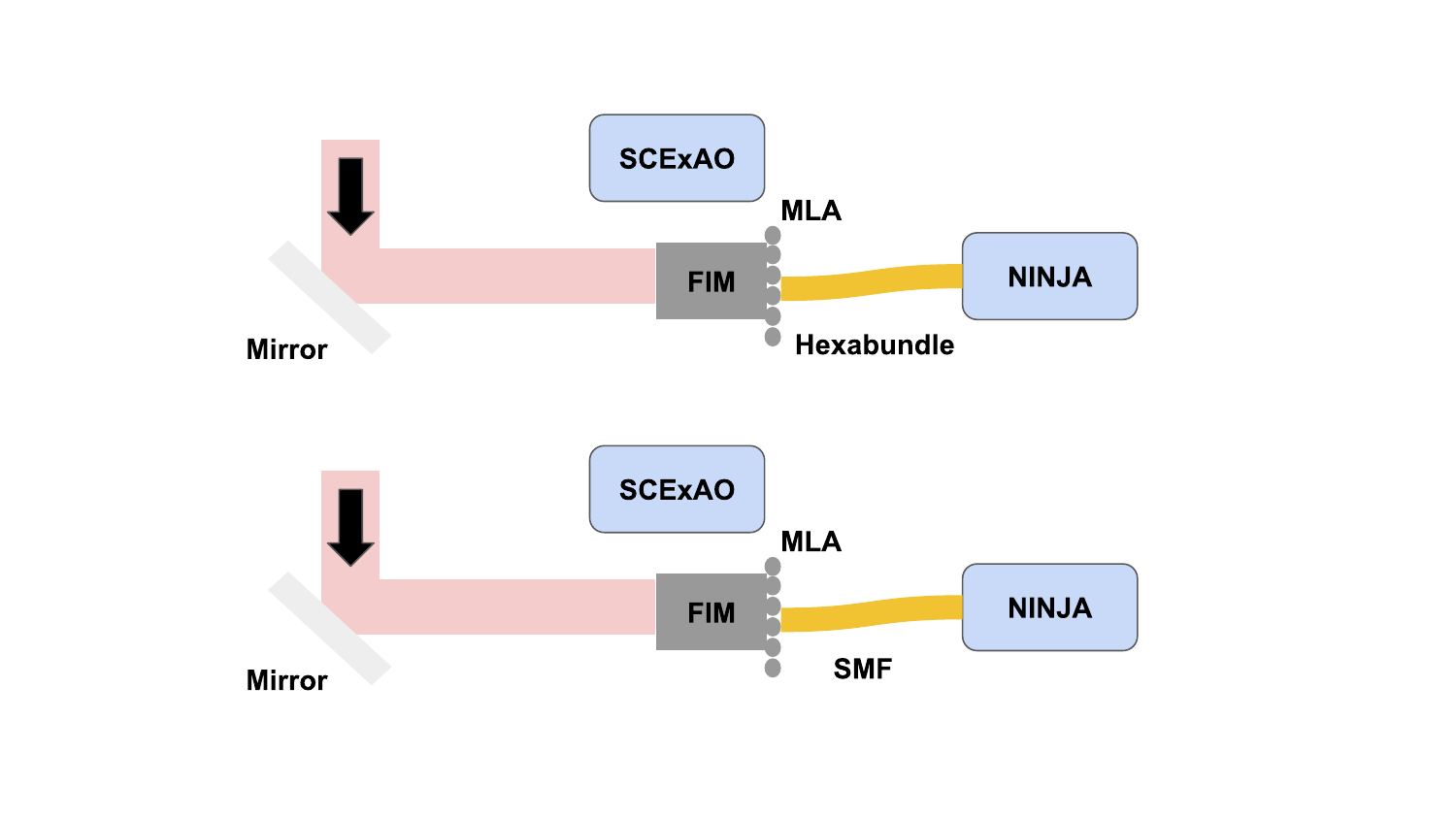}
 \caption{Diagram illustrating the two design options for Exo-NINJA. The mirror displayed in grey emulates the final off-axis parabolic (OAP) mirror in SCExAO before the Fiber Injection Module (FIM). The scenario depicted include: incorporating the microlens array (MLA) with the hexabundle (top panel), and combining the MLA with a single-mode fiber (SMF) bundle (bottom panel).}
 \label{fig:design}
\end{figure*}

Two options are considered and displayed in schematic draw in Fig. \ref{fig:design}:The top panel shows the insertion of microlenses
array (MLA) before injection into the fiber, which would change the beam speed of the incoming light onto the core of the MMF, giving an F/$\#$ that will reduce FRD losses in the fibre.
The bottom panel shows the combination of microlenses array with a fiber bundle using SMFs, with the tradeoff being a loss of throughput but FRD no longer being a concern.

We note that in all the enumerated cases above, the implementation of a fore optics is not considered as it would entail a change in the plate scale.

The optical design options will be tested at the University of Texas at San Antonio in 2024-2025 using the EGRET test bench described in the following section. Furthermore, the data reduction of Exo-NINJA will present significant challenges due to the inherent complexity of the optical design (use of MMF instead of SMF) and data acquisition processes. These complexities necessitate detailed laboratory data to develop and optimize effective data reduction techniques. The microlens arrays on the front of hexabundles is in development at Astralis-USyd and will be tested at the fibre FRD characterisation facility at The University of Sydney.

\section{Laboratory experiment:Description of the EGRET testbed}
\label{sec:lab}

\begin{figure*}
 \centering
 \includegraphics[width=0.75\textwidth]{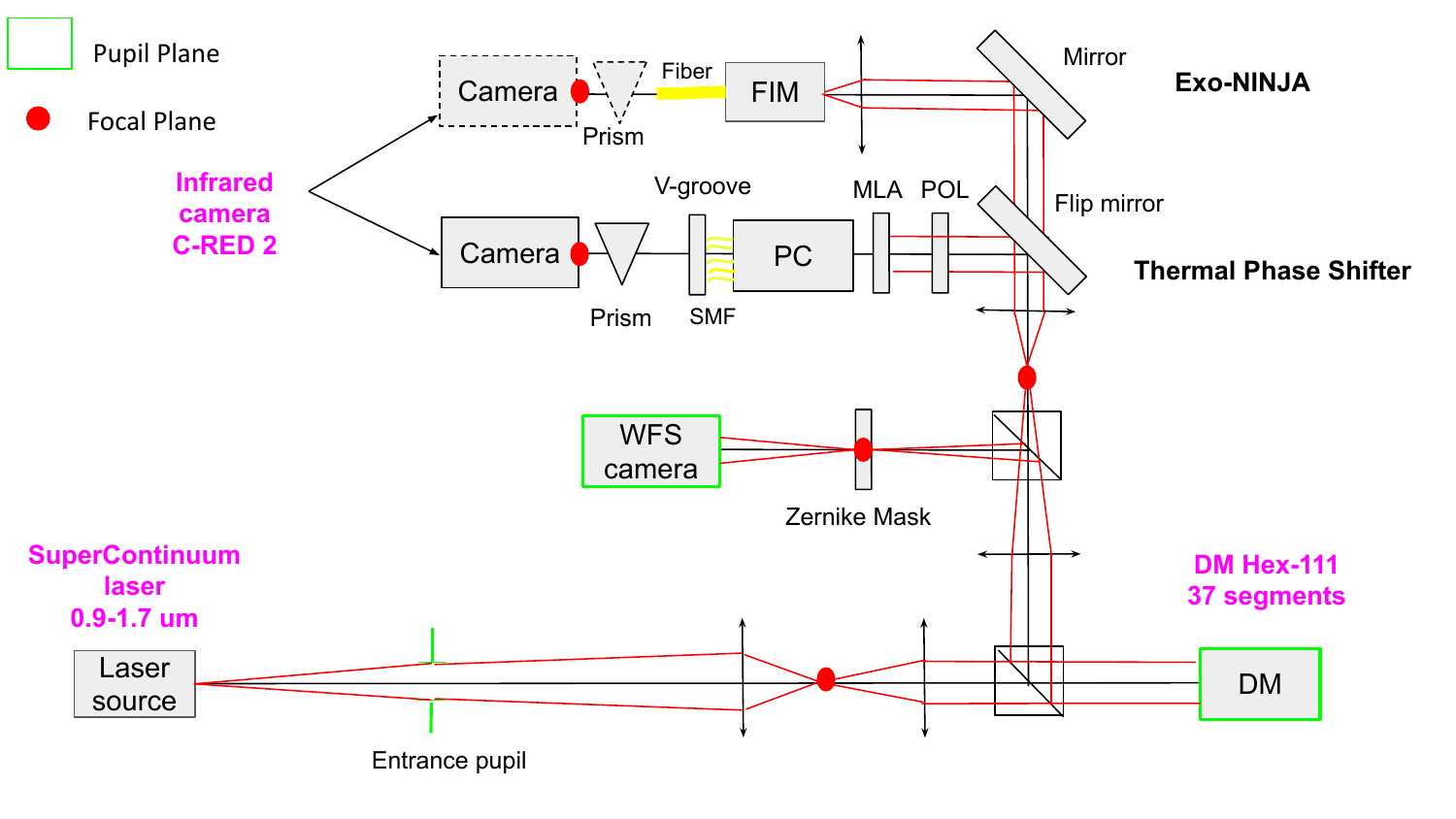}
 \caption{Schematic draw of the EGRET bench located at UTSA. The telescope simulator (lower half of the figure) includes the laser source and the WFS system. The thermal phase shifter (center) and the Exo-NINJA (top) branches are independent and only share the infrared camera, represented in straight/dashed line.}
 \label{fig:egret}
\end{figure*}

EGRET bench, displayed in Fig \ref{fig:egret}, is currently under development at UTSA, with the aim of testing innovative concepts before their implementation at Subaru. The first version of EGRET includes a telescope simulator, a Wavefront Sensor (WFS), and two independent testing branches: one dedicated to the thermal phase shifter project as part of the GLINT project \cite{Martinod2021}, and another dedicated to Exo-NINJA.

The telescope simulator consists of an Iceblink supercontinuum light source from Fyla, covering the visible (VIS) and near-infrared (NIR) spectra, with an average power exceeding 3W, pulse duration less than 10 ps, and superior stability of less than 0.5\% (standard deviation). The collimated output beam generated by the laser source passes through a reflected neutral density filter and filters mounted on an automatized wheel. Specifically, the visible spectral band is used for bench alignment, while the NIR band serves for running the tests on the bench. In the NIR band, the spectral band selected ranges from 0.9 to 1.7 nm, with a central wavelength at 1.3 nm.

The attenuated and filtered light is then injected into a Single Mode Fiber (SMF) mounted on a 3-axis mount and filtered by a circular aperture placed at the pupil plane. A set of lenses collimates the beam, directing it towards a reflective Deformable Mirror (DM) with a pupil diameter of 2.3 mm. The DM used is the gold-coated HEX-111 model from Boston Micromachine, featuring 37 hexagonal segments, each with a size of 750 $\mu$m, enabling a stroke of 3.5 $\mu$m at a 2 kHz update frequency. The light is then directed towards the WFS arm using a 50/50 beam splitter. Details of the WFS are not provided here; a feasibility study is underway to implement a Zernike wavefront sensor\cite{n2013calibration}.

A series of lenses is employed to obtain a collimated beam at the entrance of the phase-shift chip. To facilitate task alternation between the two arms, a flip mirror is integrated into the system. A mirror placed at the pupil plane emulates the last off-axis parabolic mirrors (OAP) located in SCExAO before the FIM. The light is initially injected into a simple Multi-Mode Fiber (MMF) and subsequently later into the hexabundle. The fiber bundle will be mounted on a 3-axis mount similar to the one used in the SCExAO and described previously, with a variable iris diaphragm sized allowing modulation of the F-ratio (slow, fast beam) at the bundle entrance.
The output of the fiber bundle will be positioned on a mount, allowing imaging the fiber outputs on the detector or performing injection maps. A retractable prism is added to act as a spectrograph and emulate NINJA. A translation rail is placed between Exo-NINJA and the thermal phase shifter arm to alternate camera usage. The camera is a CRED-2 from First Light with an InGaAs sensor of 640x512 pixels with a 15 $\mu$m pixel pitch. The camera, positioned at the focal plane, operates between 0.9 and 1.7 $\mu$m, with a full frame speed of 600 FPS and a readout noise at high gain of less than 30 e-.

We are testing the extent to which the F/$\#$ needs to be increased from F/3 towards the native F/14 accepted by NINJA. This requires bulk foreoptics, which must be carefully modeled along with the spectrograph optics in Zemax.


\section{Exo-NINJA Science Objectives}
\label{sec:objectives}



Exo-NINJA will be a powerful tool for studying extrasolar planets detectable by direct imaging.  Below we summarize some of the advances in exoplanet science that Exo-NINJA will enable.

\subsection{Constraining the Atmospheric Chemistry of Fully-Formed Exoplanets}

As demonstrated from first-generation instruments like VLT/SINFONI, medium-resolution spectroscopy offers a powerful method for probing exoplanet atmospheres. By applying a cross-correlation function (CCF) to theoretical templates and the planetary spectrum, molecular signatures within the atmosphere can be extracted (e.g. \cite{Hoeijmakers2018}).  Molecular signatures are distinctly visible in spatially resolved CCF images. 

This technique, known as molecular mapping, utilizes the spectral and spatial resolution of an IFU to detect molecular signatures and eliminate speckle noise, since starlight lacks these planetary molecular signatures. When combined with coronagraphy, this method achieves a contrast deep enough to recover many planets detected with advanced least-squares image processing techniques applied to traditional low-spectral-resolution IFU data.  Molecular mapping has the advantage of not requiring prior knowledge of the planet's position, making it effective for exoplanet detection. In contrast, high-resolution spectroscopy techniques like REACH\cite{kotani2020}, KPIC\cite{Delorme2021}, and HiRISE\cite{vigan2024} demand precise fiber alignment to a known planet position to efficiently inject the planet's light into the fiber.   

Current medium-resolution spectroscopic direct detections of exoplanets are limited to a handful of bright and modest-contrast companions like $\beta$ Pic b and the outer HR 8799 planets.  However, recent surveys have begun to detect a new population of planets, often fainter and closer to their host stars, thanks to improved extreme AO systems and focus on targeting stars showing dynamical evidence for a companion from astrometry \cite{Currie2023Science}.  In addition to atmospheric constraints, these planets (will) have direct dynamical mass estimates and more precisely-determined orbits thanks to their astrometric signatures.  Many of these planets -- e.g. HIP 99770 b -- should be detectable with Exo-NINJA using molecular mapping or other similar techniques.  Comparing their Exo-NINJA spectra to predictions from a suite of models will then allow us to tie planet dynamical masses to robust constraints on rotation, metallicity, C/O ratio, and other parameters not well estimated from low-resolution spectra.  

\subsection{Detecting and Measuring Gas Accretion onto Forming Planets}

Direct images of jovian protoplanets, actively-forming gas giant planets within protoplanetary disks, provide key insights into the formation and evolution of planetary systems and a context for the over 6,000 fully-formed planets detected through a variety of methods.  
Detecting gas accretion onto these protoplanets is possible through hydrogen emission, where optical wavelengths have thus far proven to be most successful: i.e. $H_{\rm \alpha}$ at 0.656 $\mu m$ \cite{Haffert2019}.  Line fluxes estimated from these detections can in turn be used to estimate planet accretion rates (e.g. \cite{Haffert2019,Hashimoto2020}); combining detections/upper limits from multiple lines may also help constrain the fractional area of the planetary surface emitting these lines (i.e. the filling factor) and the line-of-sight extinction to the protoplanet \cite{Hashimoto2020}.

However, few bona fide protoplanets have been detected thus far (e.g. PDS 70 bc; AB Aur b \cite{Keppler2018,Haffert2019,Currie2022,Currie2024}).  Many previously-identified candidate protoplanets were later shown to be disk features (e.g. LkCa 15 bcd \cite{Currie2019,Blakely2022}).   One key potential shortcoming with protoplanet detection is that embedded objects (e.g. protoplanets) may have heavily extincted optical emission, impeding the detection of their optical line emissions.  However, at least some accreting planet-mass objects show line emission at $Pa_{\rm \beta}$ (1.28 $\mu m$), which is far less sensitive to extinction\cite{Demars2023}.  High-resolution differential imaging techniques used to detect bright and modestly-embedded protoplanets in disks at $H_{\rm \alpha}$
 may then reveal more embedded protoplanets at $Pa_{\rm \beta}$ with Exo-NINJA. With the addition of Exo-NINJA, the SCExAO system will map both $H_{\rm \alpha}$ (with VAMPIRES instrument) and $Pa_{\rm \beta}$ emission.

An Exo-NINJA detection of embedded protoplanets at $Pa_{\rm \beta}$ coupled with $H_{\rm \alpha}$ upper limits or vice versa may provide key probes of \textit{how} protoplanets accrete gas.
Gas giant planets embedded in disks are expected to receive gas flows from high latitudes of the host protoplanetary disk, a process known as 'meridional flow'\cite{Kley2001}, which has been observed by ALMA\cite{Teague2019}.
 
For meridional flows, a large filling factor  (ff $>$1) would be expected, yet studies of PDS 70bc showed an ff of about 0.01\cite{Hashimoto2020}. This suggests that the emission areas are localized on the surfaces of these planets, indicating a similarity to magnetospheric accretion observed in stars.
Discovering accreting planets with a ff close to 1 would point to a new model of meridional flow accretion from the protoplanetary disk. 

\subsection{Spectro-Astrometric Detections of Protoplanets}
Detecting accreting planets at sub-diffraction limit separations is feasible by observing photocenter shifts in the 1.28 $\mu$m emission compared to the surrounding continuum. Differential measurements have already achieved precision better than 1/1000 of the point spread function (PSF) size\cite{Mendigutia2018}, equating to approximately 40 $\mu$as for the Exo-NINJA setup. This methodology enables the detection of accreting planets within sub-$\lambda$/D separations, particularly within the habitable zones of young stars. For instance, in the Taurus molecular cloud at a distance of 140 pc, where 1 AU corresponds to roughly 7 mas, this is approximately one-fifth of the $\lambda$/D limit. \\

\subsection{Enabling Future Science Observations with EXO-NINJA beyond Exoplanet Science}
Our proposed work aims to validate a technical solution for implementing a NIR fiber IFU for NINJA. This advancement will extend the instrument's capabilities beyond exoplanet research, facilitating a broad spectrum of astronomical observations. For instance, positioning the fiber bundle input directly at the telescope's Nas-IR focus will enhance throughput for observations that do not necessitate high-contrast capabilities. This setup is particularly advantageous for imaging narrow fields objects like galaxies. We plan to collaborate with the NINJA team to explore and optimize these observational options.

\section{Implementation at Subaru and status of the project}
\label{sec:status}
The NINJA spectrograph (NIR) is scheduled to be installed at Subaru in 2025. The fiber bundle has been characterized, and the final design is currently under finalization at Astralis-USyd. The fiber bundle will be sent to UTSA for the Assembly, Integration, and Testing (AIT) phase before its implementation at Subaru. The EGRET test bench, which is under development, will be used to test the Exo-NINJA concept and facilitate the acquisition of laboratory data to develop and refine a robust methodology for data analysis.

\acknowledgments 
The development of SCExAO was supported by the Japan Society for the Promotion of Science (Grant-in-Aid for Research \#23340051, \#26220704, \#23103002, \#19H00703 \& \#19H00695), the Astrobiology Center of the National Institutes of Natural Sciences, Japan, the Mt Cuba Foundation and the director's contingency fund at Subaru Telescope.

The authors wish to recognize and acknowledge the very significant cultural role and reverence that the summit of Maunakea has always had within the indigenous Hawaiian community. We are most fortunate to have the opportunity to conduct observations from this mountain.
 
\bibliography{report} 
\bibliographystyle{spiebib}

\end{document}